\documentclass[aps,prl,twocolumn,showpacs,amsmath,amssymb]{revtex4}
\usepackage{epsfig}
\usepackage{bm}

\def\a{\alpha}
\def\b{\beta}
\def\g{\gamma}
\def\G {\Gamma}
\def\d{\delta}

\def \e{\varepsilon}
\def\erg {\tau_{\rm erg}}
\def\h{\hbar}
\def\ph{\varphi}
\def\Tr{\mbox{tr}\,}
\def\S{{\cal S}}

\def\m{\mu}
\def\DD{\partial}
\def\dwell{\tau_{\rm d}}
\def \Dif {{\cal D}}
\def\Coop {{\cal C}}
\def \WS {Q}
\def \etal{{\em et al.}}
\begin{document}

\title{Mesoscopic fluctuations of nonlinear conductance of chaotic quantum dots}
\date{\today}
\author{M. L. Polianski}
\email{polian@physics.unige.ch}
\author{M. B\"uttiker}
\affiliation{D\'epartement de Physique Th\'eorique, Universit\'e de
Gen\`eve, CH-1211 Gen\`eve 4, Switzerland} 

\begin{abstract}
The nonlinear dc conductance of a two-terminal chaotic cavity is
investigated. The fluctuations of the conductance (anti)symmetric
with respect to magnetic flux inversion through multichannel
cavities are found analytically for arbitrary temperature, magnetic
field, and interaction strength. For few-channel dots the effect of
dephasing is investigated numerically. A comparison with recent
experimental data is provided.
\end{abstract}

\maketitle {\em Introduction.} Recently the non-linear dc
conductance of mesoscopic structures \cite{KL} and, more
specifically, their dependence on magnetic flux $\Phi$ has found
considerable attention \cite{wei,Zumbuhl,marlow,ensslin}.
Application of large voltages induces a rearrangement of the charge
distribution. The charge re-distribution is subject to Coulomb
interactions \cite{ChristenButtiker}. Consequently investigation of
non-linear transport reveals information on interaction parameters.
This is in marked contrast to linear transport where the dc
conductance $G_{\a\b}=dI_\a/dV_\b$ for large cavities can with high
accuracy be treated within a theory of non-interacting electrons.
However to extract this information the role of temperature and
dephasing need to be known with precision.

Interaction constants are extracted by investigating the magnetic
field symmetry of non-linear transport. Under flux $\Phi$ reversal,
in the linear regime, the Onsager-Casimir relations dictate that the
conductance matrix has the symmetry
$G_{\a\b}(\Phi)=G_{\b\a}(-\Phi)$. In particular, the conductance of
a two probe conductor is an even function of flux
\cite{Markus_1986}. However, away from equilibrium, the non-linear
conductance lacks such a symmetry. Importantly, the deviations from
Onsager symmetry are entirely due to interactions \cite{SB,SZ}.
Therefore by investigating the departure from the Onsager-Casimir
relations, information on the interaction properties can be
obtained.

Our work is motivated by very recent experiments on nonlinear
transport in various open systems: carbon nanotubes \cite{wei},
quantum dots \cite{Zumbuhl}, ballistic billiards \cite{marlow}, and
quantum rings \cite{ensslin}. Because of quantum interference, the
samples exhibit strong mesoscopic (sample-to-sample) fluctuations,
and a theory has thus to predict statistical properties. Only
recently have two theories explored such statistics for two-terminal
open samples: S\'anchez and B\"uttiker \cite{SB} considered chaotic
quantum dots in the universal regime with arbitrary interaction
strength and high magnetic fields at $T=0$, and Spivak and Zyuzin
\cite{SZ} concentrated on weakly interacting open diffusive samples
at low fields and temperatures. Although different aspects of these
theories found good agreement with experiment in quantum dots
\cite{Zumbuhl}, a more general theory that accounts for the effects
of temperature and dephasing  at arbitrary fields and interaction
strength remains to be developed. Such a theory is our main goal.
\begin{figure}[t]
\begin{center}{
\includegraphics[width=3cm]{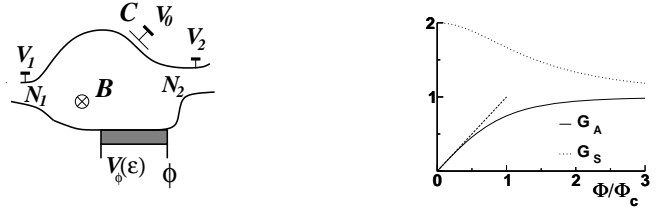}
\hfill
\includegraphics[width=3cm]{gsymasym.eps}}
\end{center}
\caption{(Left) Quantum dot with magnetic field $B$ and dc bias
voltages $V_{1,2}$ at the contacts and $V_0$ at the gates with
capacitance $C$; an additional lead $\phi$ models dephasing. (Right)
Normalized fluctuation of the (anti)symmetric component $({\cal
G}_{a}){\cal G}_{s}$ of the second order nonlinear conductance for a
coherent dot as a function of flux $\Phi/\Phi_c $. Importantly, a
crossover occurs at $\Phi_c\ll\Phi_0=eh/c$, because of the long time
an electron  spends in the dot.}\label{fig1}
\end{figure}

This Letter presents a theory of fluctuations of conductance
nonlinearity in open chaotic dots within Random Matrix Theory (RMT).
A key result of our work is illustrated in Fig. \ref{fig1}. The
(anti) symmetric parts of the non-linear conductance strongly
fluctuate from sample to sample due to quantum effects. These
fluctuations are sensitive to the flux $\Phi$ through the dot. The
fluctuations of the symmetric part ${\cal G}_{s}(\Phi)$ decrease as
the magnetic flux increases, while the anti-symmetric part ${\cal
G}_{a}(\Phi)$ has a stronger (linear \cite{SZ}) dependence at low
fields. As the flux grows, the values of fluctuations reach their
saturation values \cite{SB}. The asymptotic values are equal for
${\cal G}_{a,s}$ as expected from the linear combination of
uncorrelated contributions of random sign.

The correct definition of the crossover scale $\Phi_c$ is important
for the quantitative comparison of the theory with experiment. It
determines the slope of ${\cal G}_{a}$ at small flux $\Phi$ (see
Fig. \ref{fig1}). One might naively expect $\Phi_c \sim \Phi_0$ but
importantly we find $\Phi_c \ll \Phi_0$ in agreement with experiment
\cite{Zumbuhl}. The scale of the crossover flux $\Phi_c$ corresponds
to a flux quantum $\Phi_0=eh/c$ through a typical {\it trajectory}
of an electron and not through the {\it area} of the dot. In a
chaotic quantum dot the time $\dwell$ an electron typically spends
inside the dot is usually much larger then the ergodic time $\erg$
necessary to explore its phase space. During $\dwell/\erg$ random
attempts to explore the dot with flux $\Phi$ through its {\it area},
the flux penetrating the electronic {\it trajectory} scales with
$\Phi(\dwell/\erg)^{1/2}$. Therefore the crossover universally
occurs at $\Phi_c\sim \Phi_0(\erg/\dwell)^{1/2}\ll\Phi_0$ for
ballistic or diffusive dots (the diffusive approach \cite{SZ}
ignores $\dwell\gg\erg$, so $\Phi_c\sim \Phi_0$ of Ref. \cite{SZ}
 corresponds to the flux quantum through the {\it area} of a diffusive dot).

Below we find the fluctuations of ${\cal G}_{a}$ for arbitrary flux
$\Phi$, temperature $T$ and capacitance $C$ and compare in detail
with previously considered limits \cite{SB,SZ}. We numerically
investigate the effect of dephasing on ${\cal G}_{a}$ at high
magnetic fields, low temperatures, and strong interaction, which is
relevant for experiments. Although the dephasing diminishes the
conductance fluctuations \cite{KL,SB}, the uniform (locally weak)
dephasing is found to have a stronger effect. We conclude with a
comparison of theory and experiment.

{\em System.} The 2D quantum dot, see Fig. \ref{fig1}, is biased
with dc voltages $V_{1,2}$ at contacts with $N_{1,2}$ ballistic
channels, and by the voltage $V_0$ at the gates with capacitance
$C$. This capacitance defines the strength of the Coulomb
interaction in the dot ($C\to 0$ corresponds to strong repulsion)
\cite{ABG}. The dot is in the universal regime \cite{Beenakker},
when the Thouless energy $E_{\rm T}=\hbar/\erg$ is large, so that
the results are applicable to dots with area $A=\pi L^2$ (taken
circular), either diffusive with mean free path $l\ll L$, or
ballistic, with $l\gg L$ and chaotic classical dynamics (in the
latter case the substitution
 $l\to \pi L/4$ is used \cite{FrahmPichard}). The mean level spacing
$\Delta=2\pi\h^2/(m^*A)$ and the total number of conducting channels
$N$ together define the dwell time $\dwell=h/(N\Delta)\gg \erg$. A
dephasing with rate $\g_\ph=N_\ph\G\Delta/2\pi$ is introduced with
the dephasing probe model: a fictitious probe with $N_\ph$ channels
of transparency $\G$ is attached to the dot \cite{dJB}. We also
require that $eV\ll N\Delta$ and treat the nonlinearity only to
$(eV)^2$. Scattering is spin-independent and this spin degeneracy is
accounted for by the coefficient $\nu_s$. We use RMT for the
energy-dependent scattering matrix $\S(\e)$ and refer a reader to
reviews \cite{Beenakker,ABG} for details.

The electric potential $U$ in the dot is taken uniform. If the
screening length is much larger than the Fermi wavelength, WKB can
be applied. As a consequence, electrons with kinetic energy $\e$
have a well-defined electro-chemical potential $\tilde
\e_\a=\e-eV_\a$ in the contact $\a$ and $\tilde\e=\e-eU$ in the dot.
Therefore, transport depends on the Fermi-distributions
$f(\tilde\e_\a)$ and the scattering matrix $\S(\tilde \e)$. The
current in the contact $\a$ is $I_\a=\int d\e I_\a(\e)$ and for
$eV\ll N\Delta$ the spectral current $I_\a(\e)$ can be expanded in
powers of $eV$:
\begin{eqnarray}\label{eq:I_I}
&&I_\a (\e)=\frac{\nu_s e^2}{h}
\sum_{\d=1}^2 f(\tilde\e_\d)
\Tr \left[\openone_\a\d_{\a\d}-\openone_\d
\S^\dagger(\tilde\e)\openone_\a
\S(\tilde\e)\right]\nonumber\\
&&\approx \frac{-f'(\e)\nu_s e^2}{h}\sum_{\b}V_\b\left(
g_{\a\b}(\e)+\sum_{\g}g_{\a\b\g}(\e) eV_\g\right).
\end{eqnarray}
In Eq. (\ref{eq:I_I}) the total current $I_\a$ is expressed in terms
of the dimensionless linear conductance at energy $\e$,
$g_{\a\b}(\e)=\Tr(\openone_\a\d_{\a\b}-\openone_\b
\S^\dagger(\e)\openone_\a \S(\e))$ and the nonlinear conductance
 $g_{\a\b\g}(\e)$ related to $\DD^2 I_\a/\DD V_\b\DD V_\g$, which depends on $U$. To this
accuracy $U$ needs to be known only up to the first order
derivatives, the characteristic potentials $u_\d=\DD U/\DD V_\d$
\cite{ChristenButtiker}. The characteristic potentials $u_\d\in
(0,1)$ are found self-consistently \cite{RPA} from current
conservation and gauge-invariance requirements and expressed in
terms of the Wigner-Smith matrix $\WS=\S^\dagger \DD_\e\S/(2\pi i)$
\cite{WignerSmith}:
\begin{eqnarray}\label{eq:u}
u_\d=\frac{-\int d\e f'(\e)\Tr \WS \openone_\d}{C/e^2\nu_s -\int d\e
f'(\e)\Tr \WS},\,u_0=1-\sum_{\d=1}^2 u_\d.
\end{eqnarray}
To leading order in $N$ the mesoscopic average of Eq. (\ref{eq:u})
is flux-insensitive. However, the fluctuations of $u_\d$ are
strongly dependent on $\Phi$, and determine the asymmetry of the
non-linear conductance. These derivatives are used to express the
conductances
 $g_{\a\b\g}(\e)$:
\begin{eqnarray}\label{eq:Gabg}
g_{\a\b\g}(\e)=[\d_{\b\g} g'_{\a\b}(\e)-u_\b g'_{\a\g}(\e)-u_\g
g'_{\a\b}(\e)]/2,
\end{eqnarray}
where the prime stands for energy derivative. The matrix $\cal S$
depends on magnetic field, $\S(\Phi)=\S^T(-\Phi)$, so that
$u_0(-\Phi)=u_0(\Phi)$, but importantly $u_\d(\Phi)$ lacks such
symmetry. Therefore quite generally $g_{\a\b\g}(\Phi)\neq
g_{\a\b\g}(-\Phi)$. However some symmetries still hold for
$g_{\a\b\g}$, which is revealed in the (anti)symmetric to $\Phi\to
-\Phi$ components of conductance $({\cal G}_a) {\cal G}_s$ (in units
of inverse energy),
\begin{eqnarray}
\binom{{\cal G}_{s}}{{\cal G}_{a}}_{\a\b\g}\equiv-\int d\e
f'(\e)\frac{g_{\a\b\g}(\e,\Phi)\pm g_{\a\b\g}(\e,-\Phi)}{2},
\end{eqnarray}
which we investigate now in detail. First we derive their dependence
on temperature $T$, magnetic flux $\Phi$ and capacitance $C$ for
coherent multi-channel dots, $N\gg 1$, and later investigate
partially coherent dots at $T,C\to 0$ and high $\Phi$.

 {\em Coherent dot at arbitrary $T$, $\Phi$ and
$C$.} In a two-terminal dot without dephasing,  we use gauge
invariance and set $V_2=0$ and consider derivatives with respect to
$V_1$ only. We define ${\cal G}_{a(s)} \equiv {\cal G}_{{a(s)},111}$
and introduce a  traceless matrix $\Lambda\equiv
(N_2/N)\openone_1-(N_1/N)\openone_2$ such that
\begin{eqnarray}\label{eq:defG}
{\cal G}_{a}&=&\frac{\pi}{\Delta^2}\frac{\int\int d\e
d\e'f'(\e)f'(\e')\chi_1(\e)\chi_2(\e')}{C/(e^2\nu_s)-\int d\e
f'(\e)\Tr\WS},
\end{eqnarray}
with fluctuating $\chi_1(\e) =(\Delta/2\pi)\DD_\e \Tr\Lambda {\cal
S}^\dagger\Lambda{\cal S} $ and $\chi_2(\e) =(i\Delta/2\pi)
\Tr\Lambda[\S,\DD_\e\S^\dagger]$. The mesoscopically averaged ${\cal
G}_{a}$ vanishes, $\langle{\cal G}_a\rangle=0$, and we need to find
correlations of ${\cal G}_a$ to leading order in $N$. To this end
the products of $\S(E,\Phi)$ and $\S^\dagger(E',\Phi')$ are
 averaged, and the pair correlators, Cooperon $\Coop_{E-E'}$ and
Diffuson $\Dif_{E-E'}$, are introduced as $X_\e=(N_X-2\pi
i\e/\Delta)^{-1},X=\Coop,\Dif$, with the flux-dependent effective
number of channels $N_X$ \cite{iop}:
\begin{eqnarray}\label{eq:channels}
\binom{N_{\cal C}}{N_{\cal
D}}=N+\frac{(\Phi\pm\Phi')^2}{4\Phi_0^2}\frac{h v_F l}{L^2\Delta}.
\end{eqnarray}
The denominator of Eq. (\ref{eq:defG}) is a self-averaging quantity,
$\langle(...)^2\rangle=\langle (...)\rangle^2=(C/(C_\m\Delta))^2$,
with the electrochemical capacitance $C_\m\equiv C/(1+C\Delta/(\nu_s
e^2))$ \cite{PietMarkus}. The functions $\chi_1(\e,\Phi)$ and
$\chi_2(\e',\Phi')$ are uncorrelated, and their auto-correlations
\begin{eqnarray}\label{eq:phiphi}
\binom{\langle\chi_1\chi_1\rangle}{\langle\chi_2\chi_2\rangle}&=&
\frac {2 N_1N_2}{N^4}\binom{N_1N_2}{N^2}
\left(|\Dif_{\e-\e'}^2|\pm|\Coop_{\e-\e'}^2|\right)
\end{eqnarray}
readily allow one to find correlations of ${\cal G}_{a,s}$ at
different magnetic fields and/or temperatures. In what follows we
present the results for the variance of various experimentally
measurable quantities. Calculated at the same $T$ and $\Phi$, they
are given by very similar expressions. Most important are the
mesoscopic fluctuations of ${\cal G}_{a,s}$:
\begin{eqnarray}
\label{eq:main} \langle{\cal G}_{a}^2\rangle
&=&\left(\frac{2\pi}
{\Delta}\frac{C_\m}{C}\right)^2\frac{N_1^3N_2^3}{N^6}{\cal
F}_+{\cal F}_-,
\\
{\cal F}_\pm &=& \int \frac{d\e(\e\coth\e-1)}{\sinh^2\e}
\left(|\Dif_{2\e T}^2|\pm|\Coop_{2 \e T}^2|\right)\label{eq:Fraw}.
\end{eqnarray}
The variance $\langle{\cal G}_{s}^2\rangle$ for a realistic dot
\cite{real} with $N_1=N_2$ is given by Eq. (8) with ${\cal F}_-\to
{\cal F}_+$ . As a consequence, high magnetic field reduces the
fluctuations of ${\cal G}_{s}$ by a factor two, see Fig. \ref{fig1}.
The Eqs. (\ref{eq:main},\ref{eq:Fraw}) at high and low magnetic
fields and weak interaction are of special interest and considered
below in detail, since they allow us to compare with the results of
Refs. \cite{SB,SZ}.

 {\em Low magnetic fields.} When $N_{\cal C}\approx
N$ the magnetic field is low and can not destroy time-reversal
symmetry (TRS). In the limit $T=0$
\begin{eqnarray}\label{eq:lowBlowT}
\langle{\cal
G}_{a}^2\rangle=\left(\frac{2\pi}{\Delta}\frac{C_\m}{C}\right)^2\frac{N_1^3
N_2^3}{N^{10}} 
\left(\frac{\Phi}{\Phi_c}\right)^2,\,\,\Phi_c=\frac{\Phi_0 L
}{2\sqrt{\dwell v_F l}}.
\end{eqnarray}
At small $T\ll N\Delta/2\pi$ we have ${\cal G}_{a}^2(T)/{\cal
G}_{a}^2(0)\approx 1-8(\pi^2 T/N\Delta)^2$. For $T\gg N\Delta/2\pi$
Eq. (\ref{eq:lowBlowT}) is multiplied by $2(N\Delta/24T)^2$ and
later used to compare with experiment \cite{Zumbuhl}. We point out
that the TRS-breaking flux that destroys weak localization
 correction in open dots \cite{Chan}, obtained by other methods
\cite{Efetov95,ABG} for chaotic and disordered dots has the same
dependence \cite{Beenakker} on $\dwell$ as $\Phi_c$ in Eq.
(\ref{eq:lowBlowT}).

 For weak interaction, $C/e^2\Delta\to\infty$, we find
\begin{eqnarray}\label{eq:varGsmall}
\langle{\cal G}_{a}^2\rangle&=&\left(\frac{e^2}{2\pi
C}\frac{\dwell^2}{\hbar^2}\frac{\Phi }{\Phi_0}\right)^2\times
\left(\frac{4N_1N_2}{N^2}\right)^3\times\frac{ \dwell v_F l}{4 L^2
}.
\end{eqnarray}
We underline that Eq. (\ref{eq:varGsmall}) holds for chaotic dots
{\em independently} of the nature of scattering, diffusive or
ballistic. The first term, rewritten via $2e^2/C\to \b/\nu A$,
reproduces the result of Ref. \onlinecite{SZ}, if $\hbar/\dwell$
were substituted by the escape rate $\hbar/\erg$ of an open
diffusive sample (as is common for the crossover from ballistic to
diffusive systems).

The most important is the third term, large as $\dwell/\erg\gg 1$
for chaotic dots. It universally predicts the only relevant scale
$\Phi_c\ll\Phi_0$ rather then $\Phi_c\sim\Phi_0$ as stated in Ref.
\cite{SZ}. Data of Ref. \onlinecite{Zumbuhl}, where a nonlinearity
 with $\Phi$ sets in at flux $\Phi\ll\Phi_0$, substantiate our estimate.

{\em High magnetic fields.} When $N_{\cal C}\gg N$ the TRS is fully
broken, $\Coop/\Dif\to 0$, and at $T=0$ the functions ${\cal
F}_\pm=1/N^2$ reproduce the result of Ref. \onlinecite{SB}. At $T\gg
N\Delta/2\pi$ the asymptotes are ${\cal F}_\pm(T)\approx \Delta/(12T
N)$. For $T=0$ and weak interaction we obtain
\begin{eqnarray}\label{eq:highBlowT}
\langle {\cal G}_{a}^2\rangle &=&\left(\frac{e^2}{4
C}\frac{\dwell^2}{\hbar^2}\right)^2\times\frac{16
N_1^3N_2^3}{\pi^2N^6},
\end{eqnarray}
 where the first term reproduces the result of Ref.
\onlinecite{SZ} if $\hbar/\dwell\to \hbar/\erg$, and the second
fully accounts for a possible asymmetry in the contacts. Therefore,
at $\Phi\sim\Phi_0\gg\Phi_c$ the result (\ref{eq:highBlowT})
coincides with that of Ref. \onlinecite{SZ} up to a numerical
coefficient.

\begin{figure}[t]
\centerline{\psfig{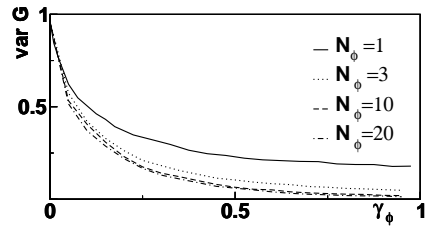}}
\caption{var ${\cal G}_{a}$, normalized by $(\pi/16\Delta)^2$, as a
function of dephasing $\g_\ph$ for $N=2,N_l=1$ for
$N_\ph=1,3,10,20$.}. \label{fig2}
\end{figure}

 {\em Partially coherent dot.} Dephasing with rate $\g_\ph$ is treated using the dephasing probe
model \cite{dJB}: the current into the dephasing probe $\ph$ is
zero, $I_\ph(\e)=0$ at every energy $\e$. The probe generates
currents in the leads $\a$ due to a "voltage" $V_\ph(\e)$ at the
probe. ($V_\ph(\e)$ defines the distribution $f(\e-eV_\ph(\e))$ at
the probe). For simplicity we take here a strongly-interacting dot,
$C=0$, and find $v_\d(\e)=\DD V_\ph(\e)/\DD V_\d=-g_{\ph
\d}(\e)/g_{\ph \ph}(\e)$ and $u_\d$:
\begin{eqnarray}
u_\d &=& \frac{\int d\e f'(\e)\Tr (\openone_\d +\openone_\ph
v_\d)\WS }{\int d\e f'(\e)\Tr \WS }. \label{eq:uderiv}
\end{eqnarray}
For a dot at $T=0$ and $\Phi\gg\Phi_c$ we numerically consider the
antisymmetric component ${\cal G}_{a}$:
\begin{eqnarray}\label{eq:Glll}
{\cal G}_{a}=\frac{u_{+}-u_{-}}{2}\left(g'_{11}+g'_{1\ph}v_+
+g'_{\ph 1}v_- +g'_{\ph\ph}v_+v_-\right),
\end{eqnarray}
with $u_\pm\equiv u_1(\pm\Phi),v_\pm\equiv v_1(\pm\Phi)$. If the
dimensionless dephasing rate $\g_\ph$, normalized by $2\pi/\Delta $,
is fixed, one can vary the transmission $\G$ together with the
 number of channels $N_\ph$ to go from uniform dephasing,
  $N_\ph\gg 1$, to non-uniform
 dephasing with a small number of perfectly conducting channels in the probe \cite{PietBeen}.
 We follow Refs. \cite{Cremers,Brouwer1995} to generate $Q$ and $\S$ for the broken
TRS for a non-ideal coupling with the probe. The results for var
${\cal G}_{a}$ as a function of $\g_\ph$ for $N=2,N_l=1$ are
presented in Fig. \ref{fig2}.
We expect that uniform dephasing damps ${\cal G}_{a}$ stronger, which is clearly seen in Fig. \ref{fig2}
 (the uniform limit $N_\ph\gg 1$ is reached at $N_\ph\sim 10$). In the limit $\g_\ph\gg N$ we have
 ${\cal G}_{a}\to 0$.
This behavior of ${\cal G}(\g_\ph)$ could also be explored in
experiments with a real additional probe (in the limit $eV,
T,\Delta\ll N\Delta$ the results of the dephasing probe \cite{dJB}
and the inelastic probe \cite{probe} models coincide).

 {\em Comparison with experiment.} Zumb\"uhl \etal \cite{Zumbuhl}
 measure the statistics of the (anti)symmetrized with respect to magnetic field $B$ conductance
$(g_{B-}) g_{B+}$ at various $V,B$. At low $eV$ they correspond to
$(2\nu_s e^3/h)V{\cal G}_{a}$  and $(\nu_s e^2/h)(g_{11}+2eV{\cal
G}_{s})$. Of particular interest are their rms $\d g_{B\pm}$ and the
coefficient $\d\alpha=\d g_{B-}/(VB)$ at $eV,B\to 0$. The
measurements of $\d\a$ are performed at $\Phi\ll\Phi_0$ and $T=4\m
eV$ for samples with $\Delta=7 \mu eV$ for $N=2,4,8$. Theory (see
\cite{SB} or Eq. (\ref{eq:channels}, \ref{eq:main}) for
$\Phi=\Phi_0$) predicts that the scaled coefficient
$\d\alpha'(N)=\d\alpha\, N^2\Phi_0 h\Delta/(2e^3 A)$ is independent
of $N$. Experiment indeed finds a reasonable agreement with this
prediction only at $N=8$ where the measured value is
$\d\alpha'(8)=1.1$ versus the predicted $\d\alpha'_{\rm th}=\pi
C_\m/2C$. Instead of a constant $\d\alpha'$ the experiment revealed
a strong dependence of $\d\a'(2)\approx 0.015,\d\alpha'(4)\approx
0.36$ on $N$. Our theory proposes an explanation of this unexpected
growth. Interestingly, it turns out that, since $T\dwell\sim \hbar$,
we can use the low-field {\it high-temperature} asymptote of Eq.
(\ref{eq:Fraw}). It provides a behavior similar to the experiment:
$\d\a'(N)=(\pi^2C_\m/24C T)(N\Delta\hbar v_{\rm F}/L)^{1/2}$. At
$\hbar v_{\rm F}/L\sim 250\mu eV$, one finds $\d\a'(N)\approx
4.5\sqrt{N} C_\m/C$ growing with $N$ but not nearly as steeply as
observed.

Experiment \cite{Zumbuhl} derives dephasing rate from weak
localization measurements, $\g_\ph=0.3$. Numerics performed at
$\Phi\gg\Phi_c$ shows that uniform dephasing $\g_\ph=0.3$ diminishes
$\d\alpha'(N)$ only by a factor 0.4, 0.65 and 0.85 for $N=2,4,8$
respectively. This is close to the result one would obtain with the
semi-empirical substitution $N\to N+\g_\ph$ in Eqs.
(\ref{eq:channels}, \ref{eq:main}), which would strongly diminish
fluctuations in the few-channel dots. As a result, assuming
non-ideal screening, $C_\m/C<1$, and strong dephasing one can fit
our results to the experimental data of Ref. \cite{Zumbuhl}. If the
experimental estimate of $\g_\ph=0.3$ and the assumption that
$C_\m/C=1$ are used, our analytical results and experimental data
disagree. Therefore both an accurate determination of the dephasing
rate and an independent measurement of the capacitance are needed.

We would like to point out that the symmetrized part of the
non-linear conductance $g_{B+}$ permits just such an independent
determination of the capacitance ratio $C_\m/C$. For symmetric ($N_1
= N_2$) coherent realistic dots \cite{real} at small $eV$ it holds
\begin{eqnarray}\label{eq:gplus}
\left(\frac{C_\m}{C}\right)^2=\frac{\mbox{var }\DD_{eV}
g_{B+}}{(\mbox{var } g_{B+})^2}\left(\frac{\nu_s
e^2}{h}\frac{\Delta}{2\pi}\right)^2.
\end{eqnarray}
at {\em arbitrary} flux $\Phi$ and temperature $T$. Since $C_\m/C$
is apriori unknown, Eq. (\ref{eq:gplus}) presents an independent way
to measure screening. Independent capacitance measurements and in
particular also experiments for asymmetric dots ($N_1 \ne N_2$), or
measurements of the nonequilibrium distribution \cite{notyet} are
needed for a more detailed comparison of theory and experiment.

{\em Conclusions.} The Onsager-Casimir relations are a cornerstone
of irreversible linear transport. We go beyond the linear regime and
consider the quantum fluctuations of transport properties due to
finite voltage and interactions by the example of open chaotic dots.
Our key universal result, confirmed by experiment \cite{Zumbuhl}, is
that the only relevant magnetic-flux scale is not the flux quantum
through a dot $\Phi_0$ but rather $\Phi_c\ll\Phi_0$ determined by a
long dwell time of electrons inside the dot. Since our theory
accounts for a wide range of parameters (temperature, flux,
dephasing, contact widths), it provides a basis for further
experimental investigation of the non-equilibrium transport.

We thank  David S\'anchez, Eugene Sukhorukov, and Dominik Zumb\"uhl
for very useful discussions, comments and data. The work was
supported by the Swiss NSF


\begin{thebibliography}{MMM}
\bibitem{KL}D.~E.~Khmel'nitskii and A.~I.~Larkin, Physica Scripta,
{\bf T14}, 4 (1986); Sov.~Phys.~JETP {\bf 64}, 1075 (1986).


\bibitem{wei}J.~Wei \etal, Phys.~Rev.~Lett. {\bf 95}, 256601 (2005).
\bibitem{Zumbuhl} D. Zumb\"uhl {\em et al}, cond-mat/0508766 (accepted in PRL).

\bibitem{marlow} C.~A.~Marlow \etal, Phys.~Rev.~Lett. {\bf 96}, 116801
(2006).

\bibitem{ensslin}R.~Leturcq \etal, Phys.~Rev.~Lett. {\bf 96}, 126801
(2006).

\bibitem{ChristenButtiker}T.~Christen and M.~B\"uttiker, Europhys. Lett. {\bf 35}, 523
(1996).


\bibitem{Markus_1986} M.~B\"uttiker, Phys.~Rev.~Lett. {\bf 57}, 1761 (1986).
\bibitem{SB}D.~S\'anchez and M.~B\"uttiker, Phys.~Rev.~Lett. {\bf 93}, 106802
(2004); Intl.~J.~Quant.~Chem. {\bf 105}, 906 (2005).

\bibitem{SZ} B.~Spivak and A.~Zyuzin, Phys.~Rev.~Lett. {\bf 93}, 226801
(2004).

\bibitem{ABG} I.~L.~Aleiner, P.~W.~Brouwer, and L.~I.~Glazman,
Phys.~Rep. {\bf 358}, 309 (2002).

\bibitem{Beenakker} C.~W.~J.~Beenakker, Rev. Mod. Phys. {\bf 69}, 731 (1997).

\bibitem{FrahmPichard} K.~Frahm and J-L.~Pichard, J. Phys.
(France) I {\bf 5}, 847 (1995); S.~Adam \etal, Phys.~Rev.~B {\bf
66}, 195412 (2002).


\bibitem{dJB} M.~J.~M. de Jong and C.~W.~J.~Beenakker, Physica A {\bf 230},
219 (1996).

\bibitem{RPA} M. B\"uttiker, A. Pr\^etre, and H. Thomas, Phys. Rev. Lett. {\bf 70}, 4114
(1993).

\bibitem{WignerSmith} E. P. Wigner, Phys. Rev. {\bf 98}, 145 (1955); F. T. Smith,
Phys. Rev. {\bf 118}, 349 (1960).

\bibitem{iop} M.~L.~Polianski and P.~W.~Brouwer, J.~Phys.~A: Math.~Gen.
{\bf 36}, 3215 (2003).



\bibitem{PietMarkus} P.~W.~Brouwer and M.~B\"uttiker, Europhys.\ Lett. {\bf 37},
441 (1997).

\bibitem{real} Here $e^2/C\gg\mbox{ max }(N\Delta,(k_{\rm B} T N\Delta
)^{1/2})$.

\bibitem{Chan} I.~H.~Chan \etal, Phys.~Rev.~Lett. {\bf 74},3876
(1995).

\bibitem{Efetov95} K.~B.~Efetov, Phys.~Rev.~Lett. {\bf 74}, 2299
(1995).



\bibitem{PietBeen} P.~W.~Brouwer and C.~W.~J.~Beenakker, Phys.~Rev.~B {\bf 55}, 4695
(1997).



\bibitem{Cremers} J.~N.~H.~J.~Cremers and P.~W.~Brouwer, Phys.~Rev.~B {\bf 65}, 115333
(2002).
\bibitem{Brouwer1995} P.~W.~Brouwer, Phys.~Rev.~B {\bf 51}, 16878
(1995).

\bibitem{probe}M.~B\"uttiker, IBM~J.~Res.~Dev. {\bf 32}, 63
(1988).

\bibitem{notyet} D.~Zumb\"uhl \etal (unpublished).

\end{thebibliography}
\end{document}